\documentclass{PoS}

\def\beq{\begin{equation}}
\def\eeq{\end{equation}}
\def\bea{\begin{eqnarray}}
\def\eea{\end{eqnarray}}

\def\ep{\epsilon}

\def\nn{\nonumber}
\def\Eq#1{Eq.~(\ref{#1})}

\def\td#1{\tilde{\delta}\left(#1\right)}

\newcommand{\la}{\langle}
\newcommand{\ra}{\rangle}

\def\qb{\mathbf{q}}

\title{
\vspace*{-2.5cm}
\begin{minipage}{\textwidth}
{\normalfont\small IFIC/15-71, LPN15-027
\hspace{\fill} October 2015
}\\
\end{minipage}\\[60pt]
  From dimensional regularization to NLO computations in four dimensions}

\ShortTitle{From DREG to NLO computations in 4D}

\author{\speaker{Germ\'an F. R. Sborlini}$^{a,b}$, Roger Hern\'andez-Pinto$^a$ and Germ\'an Rodrigo$^a$\\\\
        $^a$Instituto de F\'{\i}sica Corpuscular, Universitat de Val\`{e}ncia -- 
Consejo Superior de Investigaciones Cient\'{\i}ficas, 
Parc Cient\'{\i}fic, E-46980 Paterna, Valencia, Spain. \\\
				$^b$Departamento de F\'\i sica and IFIBA, FCEyN, Universidad de Buenos Aires, 
(1428) Pabell\'on 1 Ciudad Universitaria, Capital Federal, Argentina. \\\\
        E-mail: \email{german.sborlini@ific.uv.es, rogerjose.hernandez@ific.uv.es, german.rodrigo@csic.es}}

\abstract{Loop-tree duality (LTD) allows to express virtual contributions in terms of phase-space integrals, thus leading to a direct mapping with real radiation terms. We review the basis of the method and describe its application to regularize Feynman integrals. Performing an integrand-level combination of real and virtual terms, we show how to recover physical results by simply taking the four-dimensional limit of $d$-dimensional expressions. Moreover, this method provides a natural physical interpretation of infrared singularities, their origin and the way that they cancel in the complete computation.}

\FullConference{The European Physical Society Conference on High Energy Physics\\
		22--29 July 2015\\
		Vienna, Austria}

\begin{document}

\section{Introduction and motivation}
\label{sec:introduction}
Obtaining physical results from quantum field theories (QFT) involves dealing with ill-defined expressions in intermediate steps of the calculation and a regularization method is required to explicitly show these problems. The customary method is dimensional regularization (DREG) \cite{Bollini:1972ui, 'tHooft:1972fi, Cicuta:1972jf, Ashmore:1972uj}, which consists in performing an analytical extension from $d=4$ to $d=4-2\epsilon$ space-time dimensions: singularities appear as poles in $\epsilon=0$. Those poles originated in the ultraviolet (UV) region can be cancelled by the application of a proper renormalization scheme \cite{Hernandez-Pinto:2015ysa}. On the other hand, when computing \emph{infrared-safe observables}, KLN theorem \cite{Kinoshita:1962ur} guarantees the cancellation of infrared (IR) divergences after combining real and virtual corrections. This is related with the fact that the divergent structure of loop amplitudes is deeply connected with the physical divergences developed by real radiation processes integrated in singular regions of the phase-space (PS). 

This knowledge has been exploited by many computational methods, specially those relying in subtraction techniques \cite{Catani:1996vz,Catani:1996jh,Frixione:1995ms,GehrmannDeRidder:2005cm,Catani:2007vq,DelDuca:2015zqa,Czakon:2010td,Boughezal:2015dva,Gaunt:2015pea}. The standard approach consists in adding to the real radiation contribution suitable counter-terms that mimic the singular behavior in the IR limits, cancelling the corresponding soft and collinear divergences. The same quantity, integrated over the PS of the extra radiation, is subtracted back from the virtual corrections. For instance, at next-to-leading order (NLO) this reads
\bea
\sigma^{\rm NLO} &=& \int_m \left[d\sigma^{\rm 1-loop}-\int_1 dA\right] \, + \int_{m+1} \left[d\sigma^{\rm real} + dA\right] \, ,
\eea
where $dA$ is the local form of the counter-term in the $m+1$-particle PS. This formula is naturally extended to higher-orders by adding all the possible combination of loop and extra-radiation contributions. Notice that a key point of the procedure is the possibility to perform an analytical integration of the counter-term $dA$, since we need to combine it with virtual terms which involve less particles in the final state. However, increasing the number of legs and loops leads to cumbersome counter-terms which reduces the efficiency of this technique. Moreover, intermediate steps must be carried out in $d$-dimensions to handle IR divergences properly, because $\epsilon$-poles are explicitly present in $d\sigma^{\rm 1-loop}$ (originated in \emph{loop} integration) whilst their real counter-part appear after integrating $d\sigma^{\rm real}$ (free of $\epsilon$-poles) over the $m+1$ PS.

The purpose of this presentation is to explain an alternative approach based in the loop-tree duality (LTD). As stated in Ref. \cite{Catani:2008xa}, loop integrals or scattering amplitudes in relativist, local and unitary QFT are related with PS integrals of tree-level objects with modified propagators, which are called \textit{dual propagators}. For the sake of simplicity, we restrict the discussion to one-loop level, although these ideas are naturally extended to higher-loops \cite{Catani:2008xa,Bierenbaum:2010cy,Bierenbaum:2012th,Buchta:2014dfa}. So, given an $N$-leg scalar one-loop integral its dual representation is obtained as the sum of $N$ dual integrals associated with each possible one-cut, i.e.
\bea
L^{(1)}(p_1, \dots, p_N) 
&=& - \sum_{i\in \alpha_1} \, \int_{\ell} \; \td{q_i} \,
\prod_{j \in \alpha_1, \, j\neq i} \,G_D(q_i;q_j)~, \nn \\
\label{oneloopduality}
\eea 
where $G_D(q_i;q_j) = (q_j^2 -m_j^2 - i0 \, \eta \cdot k_{ji})^{-1}$ are dual propagators, with $i,j \in \alpha_1 = \{1,2,\ldots N\}$. The momenta of the internal lines are denoted $q_{i,\mu} = (q_{i,0},\mathbf{q}_i)$, where $q_{i,0}$ is the energy and $\qb_{i}$ are the spacial components. They are defined as $q_{i} = \ell + k_i$ with $\ell$ the loop momentum and $k_{i} = p_{1} + \ldots + p_{i}$. The four-momenta of the external legs are $p_{i}$, which are taken as 
outgoing, with $k_{N} = 0$ by momentum conservation. Since Cauchy's residue theorem lays behind the derivation of this formula, counter-clockwise orientation must be used to describe internal line momenta flow. Besides that, $\eta$ is an arbitrary space-like or light-like vector with positive energy component; it is crucial for the definition of the dual prescription which takes care of the multiple-cut contributions involved in the Feynman's tree theorem (FTT) \cite{Feynman:1963ax,Feynman:1972mt} and allows to exactly recover the discontinuity structure of virtual amplitudes.

From \Eq{oneloopduality}, notice that the integration measure becomes
\beq
\int_{\ell} \, \td{q_i}\bullet = - \imath \mu^{4-d} \int \frac{d^d \ell}{(2\pi)^{d}} \td{q_i} \bullet~, \ \ \ \ \ \ \td{q_i} \equiv 2 \pi \, \imath \, \theta(q_{i,0}) \, \delta(q_i^2-m_i^2) \, ,
\eeq
where the delta distribution sets internal lines on-shell and forces them to have positive energy ($q_{i,0}>0$). So, LTD converts the usual $d$-dimensional loop measure into a $(d-1)$-dimensional integration over the forward on-shell hyperboloid associated with the equation $G_F(q_i)^{-1}=(q_i^2-m_i^2+\imath 0)=0$, which resembles a typical real-radiation PS measure without imposing momentum conservation. It is worth appreciating that on-shell hyperboloids degenerate to light-cones (LC) for massless propagators, and it is strongly related with the presence of IR singularities.

In the following we explore the consequences of the application of LTD to some simple processes. A more detailed version of this presentation can be found in Ref. \cite{Hernandez-Pinto:2015ysa}. In Sec. \ref{sec:IRsingularities}, we analyse the origin of IR divergences in a scalar triangle Feynman integral, and we prove that they are contained in a compact region of the integration domain. Then, in Sec. \ref{sec:realvirtual}, the procedure to combine real and \emph{dual} contributions is explained, focusing in the relations among different kinematical variables. Finally, the conclusions and outlook are presented in Sec. \ref{sec:conclusions}.

\section{IR singularities through LTD}
\label{sec:IRsingularities}
Let's start by considering a scalar triangle Feynman integral in the time-like (TL) region. The process is represented by the kinematical configuration $p_3 \to p_1 + p_2$, with $p_1^2=0=p_2^2$ and $p_3^2=s_{12}>0$. If $\ell$ denotes the loop momentum, then $q_1 = \ell + p_1$, $q_2 = \ell + p_{12}$ and $q_3 = \ell$ describe the momenta at the internal lines. The application of LTD leads to 
\bea
L^{(1)}(p_1,p_2,-p_3) &=& \sum_{i=1}^3 I_{i}~ = \frac{c_{\Gamma}}{s_{12}\,\epsilon^2} \left(\frac{-s_{12}-\imath 0}{\mu^2}\right)^{-\epsilon} \, ,
\label{triangulo}
\eea
with 
\bea
I_1 &=& \frac{1}{s_{12}} \, \int_{0}^{\infty} d[\xi_{1,0}] \, \int_{0}^{1} d[v_1] \,  
\xi_{1,0}^{-1} \, (v_1 (1-v_1))^{-1}~, 
\label{dualintegrals1}
\\ I_2 &=& \frac{1}{s_{12}} \, \int_{0}^{\infty} d[\xi_{2,0}] \, \int_{0}^{1} d[v_2] \,  
\frac{(1-v_2)^{-1}}{1 - \xi_{2,0} + \imath 0}~, 
\label{dualintegrals2}
\\ I_3 &=& - \frac{1}{s_{12}} \, \int_{0}^{\infty} d[\xi_{3,0}] \, \int_{0}^{1} [v_3] \,  
\frac{v_3^{-1}}{1 + \xi_{3,0}}~,
\label{dualintegrals3}
\eea
where we defined the $d$-dimensional integration measures as 
\beq
d[\xi_{i,0}] = \frac{\mu^{2\ep} \, (4\pi)^{\ep-2}}{\Gamma(1-\ep)} \,
s_{12}^{-\ep} \, \xi_{i,0}^{-2\ep} \, d\xi_{i,0}~, \ \ \ \ d[v_i] = (v_i(1-v_i))^{-\ep} \, dv_i~.
\eeq
Notice that Eqs. (\ref{dualintegrals1})-(\ref{dualintegrals3}) are valid in the center-of-mass frame, with $\mathbf{p}_1$ ($\mathbf{p}_2$) along the positive (negative) $z$-axis, such that $2 q_i\cdot p_1 / s_{12} = \xi_{i,0}\, v_i$ and $2 q_i\cdot p_2 / s_{12} = \xi_{i,0}\, (1-v_i)$. The dual prescription plays a crucial role in $I_2$, since the denominator vanishes inside the integration domain. This is related with the presence of a threshold singularity; when $\xi_{2,0}=1$, then two internal lines become simultaneously on-shell and the diagram can be split into two physical tree-level contributions \cite{Rodrigo:prep}. 

\begin{figure}[ht]
\begin{center}
\includegraphics[width=5.5cm]{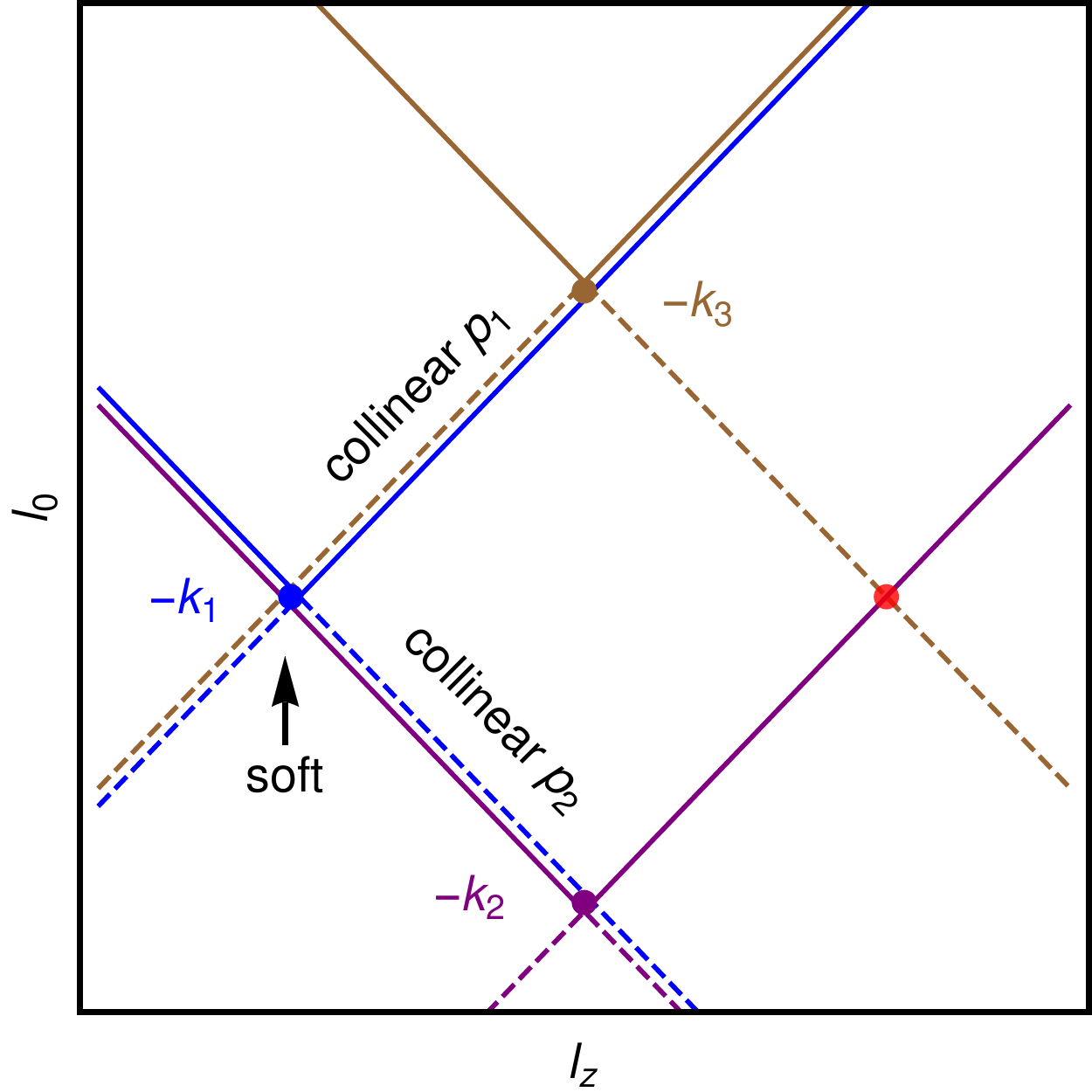}
\includegraphics[width=5.5cm]{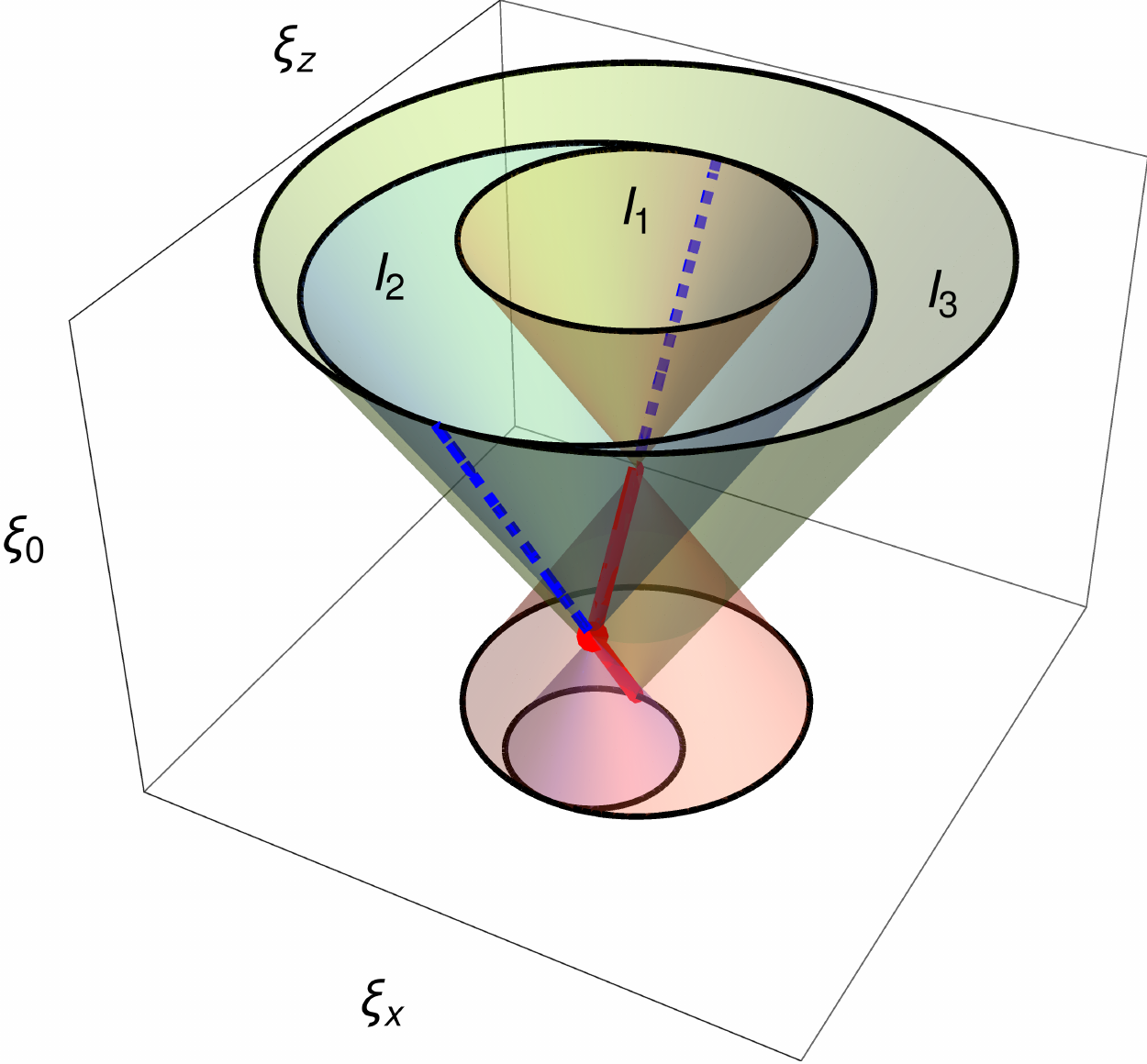}
\caption{Location of threshold and IR singularities in the $(\ell_0,\ell_z)$ space (left); forward-forward singularities cancel among dual contributions whilst forward-backward intersections lead to IR and threshold singularities. Three-dimensional representation of the light-cones and their intersections (right). 
\label{fig:IRsingularities}}
\end{center}
\end{figure}

On the other hand, the dual representation allows to identify the origin of the singularities present in the loop integrand. In Fig. \ref{fig:IRsingularities}, we show the light-cones associated with the solutions of $G_F(q_i)^{-1}=0$ in the loop-momentum space; the integration domain of the dual integrals is restricted to the forward LC (solid lines). As explained in Ref. \cite{Buchta:2014dfa}, intersection among LCs give rise to singularities in the dual integrands that can be related with those present in the original loop integrands. In this example, the punctual intersection among the three LC produces the soft singularity, that originates a double $\epsilon$-pole in the IR region of $I_i$. Collinear singularities are described by forward-backward intersections. If we interpret the cut-line as a physical particle with momentum $q_i$, then the intersection of the forward LC for $I_1$ and the backward LC for $I_3$ corresponds to the configuration $q_1 \parallel p_1$. This is associated with the presence of a factor $v_i^{-1}$ in $I_1$ and $I_3$. An analogous interpretation can be used to describe the collinear singularity when $q_1 \parallel p_2$, which involves contributions coming from $I_1$ and $I_2$. The integrable threshold-singularity is associated with the intersection of the forward LC for $I_2$ and the backward LC for $I_3$. In that case, $q_2^2=0=q_3^2$ and all the on-shell lines have positive energy. Finally, forward-forward intersections also produces divergences, but they cancel when combining dual integrands.

This graphical analysis motivates the definition of new integrals which allows to isolate IR singularities in a compact region of the loop momentum space. Introducing a technical cut $w>0$ to properly treat the threshold region \cite{Rodrigo:prep}, we define
\beq
I^{\rm IR} = I_1(\xi_{1,0}\leq w) + I_1(w\leq\xi_{1,0}\leq 1;v_1 \leq 1/2) + I_2(\xi_{2,0}\leq 1+w;1/2 \leq v_2) \, ,
\eeq
and combine the remaining contributions inside the so-called forward and backward integrals, i.e.
\bea
I^{\rm (b)} &=& I_1(w \leq \xi_{1,0};1/2 \leq v_1) +I_2(1+w \leq \xi_{2,0};1/2 \leq v_2) + I_3(1/2 \leq v_3) \, ,
\\  I^{\rm (f)} &=& I_1(1 \leq \xi_{1,0};v_1 \leq 1/2) +I_2(v_2 \leq 1/2) + I_3(v_2 \leq 1/2) \, ,
\eea
such that $L^{(1)}(p_1,p_2,-p_3) = I^{\rm IR} + I^{\rm (b)} + I^{\rm (f)}$. The IR-divergent structure of $I^{\rm IR}$ agrees with that of the scalar triangle, given in \Eq{triangulo}. On the other hand, $I^{\rm (f)}$ and $I^{\rm (b)}$ are finite in the limit $\epsilon \to 0$, and can be computed in four-dimensions. For this purpose, it is requested to unify the coordinate system (expressing $(\xi_{1,0},v_1)$ in terms of $(\xi_{3,0},v_3)$) in order to achieve an exact matching of singularities at integrand level. After this change of variables, we can take the limit $\epsilon \to 0$ at integrand level and recover the expected result.

\section{Real+virtual mapping}
\label{sec:realvirtual}
At this point, we know that IR singularities associated with the scalar triangle are isolated in a compact region of the integration domain. So, let's consider a massless scalar QFT and the corresponding three-point one-loop amplitude. We assume that it is proportional to the triangle, i.e. $| {\cal M}^{(1)} (p_1, p_2; p_3)\ra = -\imath \, g^3 \, L^{(1)}(p_1, p_2, -p_3)$, where $g$ is an arbitrary coupling. According to the KLN theorem, the NLO correction to the total-inclusive cross section is IR-safe. This implies that we can cancel all IR-singularities present in the scalar triangle by adding the real contribution.

The key feature of the LTD approach is that real and virtual contributions can be combined at integrand level, in spite of being described by different kinematics ($1 \to 2$ for LO and one-loop terms, whilst real radiation is associated to $1 \to 3$ processes). So, we introduce a momenta mapping to relate $p_3 \to p_1 + p_2$ and loop-momentum $\ell$ (virtual kinematics) with $p_3 \to p_1'+p_2'+p_r'$ (real kinematics). To achieve a fully local cancellation of IR singularities, we split the real-radiation phase-space in two disjoint regions, according to
\beq
{\rm PS}^{1 \to 3} = \{s_{1r}'<s_{2r}'\} \bigcup \{s_{2r}'<s_{1r}'\} = {\cal R}_1 \bigcup {\cal R}_2 ,
\eeq
which are characterized by containing only one collinear configuration ($1 \parallel r$ in ${\cal R}_1$ and $2 \parallel r$ in ${\cal R}_2$). Using the dimension-less variables $y_{ir}'= s_{ir}'/s_{12}$ for the real process and $(\xi_{i,0},v_i)$ for the dual part, we can use the mapping
\beq
y_{1r}' = \frac{v_1 \, \xi_{1,0}}{1-(1-v_1) \, \xi_{1,0}}~, \quad 
y_{2r}' = \frac{(1-v_1) \, (1-\xi_{1,0}) \, \xi_{1,0}}{1-(1-v_1) \, 
\xi_{1,0}}~, \quad y_{12}' = 1-\xi_{1,0} \, ,
\eeq
in ${\cal R}_1$; a completely analogous map is defined in ${\cal R}_2$\footnote{A detailed discussion about these formulae can be found in Refs. \cite{Hernandez-Pinto:2015ysa, Buchta:2015xda, Buchta:2015wna}. The formal derivation and its extension to multi-leg processes will be presented in Ref. \cite{Rodrigo:prep}.}. Then, we introduce the real cross-sections
\beq
\widetilde \sigma_{R,i}^{(1)} = 2 {\rm Re} \int d\Phi_{1\to 3} 
\, \la{\cal M}^{(0)}_{2r} |{\cal M}^{(0)}_{1r}\ra \, \theta(y_{jr}'-y_{ir}')~,  
\eeq
that fulfil $\widetilde \sigma_{1,R}+\widetilde \sigma_{2,R}=\sigma^{\rm real}$ and can be expressed in terms of $(\xi_{1,0},v_1)$ and $(\xi_{2,0},v_2)$, respectively. On the other hand, direct application of LTD to the virtual amplitude leads to \emph{dual cross-sections}, given by
\beq
\widetilde \sigma_{V,i}^{(1)} = 2 {\rm Re} \int d\Phi_{1\to 2} \, 
\la{\cal M}^{(0)} |{\cal M}_i^{(1)}\ra \, \theta(y_{jr}'-y_{ir}')~, 
\label{eq:sigmas}
\eeq
whilst the remainders are used to define $\widetilde \sigma^{(f)}$ and $\widetilde \sigma^{(b)}$, which can be implemented in four-dimensions as described in Sec. \ref{sec:IRsingularities}. Finally, we combine real and dual cross-sections to obtain
\bea
\nn \widetilde  \sigma_1^{(1)} &=& \widetilde \sigma_{V,1}^{(1)} + \widetilde \sigma_{R,1}^{(1)} = 
2 g^2 \, \sigma^{(0)} \, \int d[\xi_{1,0}] \, d[v_1] \, \theta(1-2 v_1) \,
\theta\left(\frac{1-2v_1}{1-v_1}-\xi_{1,0}\right)  \, 
\\ &\times& \xi_{1,0}^{-1} (v_1 (1-v_1))^{-1} \, \left[  
\left(\frac{1-\xi_{1,0}}{1-(1-v_1)\, \xi_{1,0}} \right)^{-2\ep} - 1 \right]\, \equiv 0 + {\cal O}(\ep) \, ,
\label{eq:sigma1}
\\ \nn \widetilde \sigma_2^{(1)} &=& \widetilde \sigma_{V,2}^{(1)} + \widetilde \sigma_{R,2}^{(1)} =2 g^2\, \sigma^{(0)} \, \int d\xi_{2,0} \, dv_2 \, \theta\left(\frac{1 - \sqrt{1-v_2}}{v_2} - \xi_{2,0}\right) 
 \, 
\\ &\times& (1-v_2)^{-1} \bigg[\frac{1}{1-v_2 \xi_{2,0}}
- \frac{1}{1-\xi_{2,0}+\imath 0} - \imath \pi \delta(1-\xi_{2,0})\bigg]\, + {\cal O}(\ep)~,
\label{eq:sigma2}
\eea 
where we add the Cutkowsky component to get rid of the imaginary part. It is crucial to appreciate that $\sigma_1$ are represented by integrable functions in four-dimensions; i.e. we can take the limit $\epsilon \to 0$ at integrand level and obtain the finite correction to the total cross section.

\section{Conclusions and outlook}
\label{sec:conclusions}
In this presentation we explored the consequences of the LTD, focusing in the treatment of physical singularities. In particular, the introduction of dual integrals leads to a natural interpretation of the IR structure of the scalar triangle, which allowed us to prove that they are originated inside a compact region ${\cal R}_{\rm IR}$ of the integration domain. The importance of this fact can be fully appreciated when we combine real and virtual contributions. 

By the introduction of a proper real-radiation PS separation and defining suitable momentum mappings in each region, we expressed real and virtual contributions using the same integration variables. We rewrited $\sigma^{\rm real}$ using $\widetilde \sigma_{R,i}^{(1)}$, which naturally involve integrands with compact support. Then, we properly decomposed ${\cal R}_{\rm IR}$ to define dual cross-sections $\sum \, \widetilde \sigma_{V,i}^{(1)}$ that contain all the IR-poles associated with the virtual contribution. Since the result must be finite according to the KLN theorem and the mapping guarantees the same behaviour of real and dual integrands in the IR-limits, the combined cross-sections can be computed by taking the limit $\epsilon \to 0$ at integrand level. This is a crucial advantage of this method compared with the traditional subtraction approach, because we avoid the introduction of IR counter-terms.

The possibility of performing purely four-dimensional implementations could lead to major improvements in the computation of higher-order corrections in QFT, besides allowing a better understanding of the mathematical structures behind scattering amplitudes.

\section*{Acknowledgements}
This work has been supported by the Research Executive Agency (REA) under the Grant Agreement No. PITN-GA-2010-264564 (LHCPhenoNet), by CONICET Argentina, by the Spanish Government and ERDF funds from the European Commission (Grants No. FPA2014-53631-C2-1-P, FPA2011-23778, CSD2007-00042 Consolider Project CPAN) and by Generalitat Valenciana under Grant No. PROMETEOII/2013/007.


\end{document}